\documentclass[12pt]{article}%
\usepackage{citesort}
\usepackage{amsmath}
\usepackage{graphicx}%
\usepackage{amsfonts}%
\usepackage{amssymb}
\def\ba{\begin{eqnarray}}
\def\ea{\end{eqnarray}}

\begin{document}

\title{ Wounded quarks and diquarks \\in high energy collisions}
\author{A. Bialas and A. Bzdak\thanks{Fellow of the Polish Science Foundation (FNP)
scholarship for the year 2006-2007}\\M. Smoluchowski Institute of Physics \\Jagellonian University, Cracow\footnote{Address: Reymonta 4, 30-059 Krakow,
Poland; e-mail: bialas@th.if.uj.edu.pl}}
\maketitle

\begin{abstract}
Particle production in $Au-Au$, $Cu-Cu$, $d-Au$ and $p-p$ collisions at 200
GeV c.m. energy are analyzed in the wounded quark-diquark model. Existing data
are well reproduced. Emission functions of wounded and unwounded constituents
are determined. Implications for the collective evolution of the system are discussed.

\end{abstract}

\section{Introduction}

It is now well established that the idea of a "wounded" source of particles
\cite{bbc} turned out rather useful in description of particle production from
nuclear targets \cite{busza}. It seems therefore interesting to verify to what
extent the recent high energy data confirm this hypothesis and what are its
consequences for the evolution of the system created in heavy ion collisions
at high energy. This is the subject of the present paper.

A wounded source, by definition, emits a fixed density of particles,
independently of the number of collisions it underwent inside the nucleus. To
understand the physical meaning of this concept, let us recall the original
argument leading to this idea.

The idea can be understood if one observes that the process of particle
production is not instantaneous. This was first noted (and the consequences
for scattering at high energies derived) by Landau and Pomeranchuk
\cite{lanpom}. A simplified version of the argument can be presented as follows.

Consider a particle created in a high-energy collision. In the reference frame
where the longitudinal momentum of this particle vanishes, the minimal time
necessary for its creation is $t_{0}\geq1/m_{t}$ where $m_{t}=\sqrt
{m^{2}+p_{t}^{2}}$ is its energy. Consider now the ''laboratory'' frame where
the target nucleus is at rest. In this frame the particle in question acquires
some longitudinal momentum, the time is multiplied by the Lorentz factor, and
we have%
\begin{equation}
t=\gamma t_{0}\geq\frac{E}{m_{t}^{2}}=\frac{\cosh y_{lab}}{m_{t}},
\end{equation}
where $E$ is the energy of the particle. Consequently, the uncertainty of the
distance from the collision point to that at which the particle is created
(i.e. the resolving power of the longitudinal distance) is
\begin{equation}
l=vt\geq\frac{\sinh y_{lab}}{m_{t}}. \label{l}%
\end{equation}

When this distance is longer than the size of the nucleus, $Z(b)$, at a given
impact parameter i.e. when the rapidity of the produced particle is large
enough, the particle cannot resolve individual collisions and therefore it is
natural to admit that its creation may be insensitive to the number of
collisions of the source. This is the origin of the idea of wounded sources
and it is clear that it makes sense only when one considers production of
particles with the laboratory rapidity exceeding that determined by the
condition $l\geq Z(b)$. But at small laboratory rapidities particle production
from a fast-moving source is anyway expected to be small (if not negligible).
Therefore the concept of the wounded source can be applied in practice in the
whole rapidity region.

It was recently shown \cite{bb-AA} that the idea of wounded constituents of
the nucleon can successfully explain the centrality dependence of particle
production at $\eta=0$ in $Au-Au$ collisions at RHIC \cite{phob-AuAu}.
Assuming that high energy interactions of nucleons are dominated by
independent interactions of its two constituents, a quark and a diquark (each
one producing the same particle density), it was possible to explain the small
momentum transfer elastic $pp$ and $\pi p$ scattering \cite{bb-pp} as well as
energy and centrality dependence of particle production in $Au-Au$ collisions
at the same time\footnote{Other efforts in this direction can be found in
\cite{other}.}.

Encouraged by these results, in the present paper we extend the analysis to
rapidities in a wide range beyond the central region using data on particle
production in $p-p,$ $d-Au$, $Au-Au$ and $Cu-Cu$ collisions at $200$ GeV c.m. energy.

Our main conclusion is that the data are in reasonable agreement with the
wounded quark-diquark model in a wide range of rapidities excluding, however,
the narrow target and projectile fragmentation regions. In the fragmentation
regions two additional contributions could be identified:

(i) First, it is necessary to account for the decay products of the
constituents which did not interact (''unwounded'') but belong to a wounded
nucleon. Since such constituents are coloured, they have to decay into
observable particles.

(ii) One also has to take into account particle production by secondary
interactions of particles produced by the projectile inside the nucleus.
Since, as already explained, the fast particles (in the rest frame of the
nucleus) are produced outside the nucleus, the secondary interaction may only
produce particles which are slow in the rest frame of the nucleus.

In the next section we present a general formulation of the model. The $d-Au$
collisions are discussed in Section $3$ and the nucleus-nucleus collisions in
Section $4$. Our conclusions are listed in the last section where also some
comments are included.

\section{General formulation of the model}

In the central rapidity region the model predicts that the density of
particles produced in collision of nucleus $A$ with nucleus $B$ is given by
\cite{bb-AA}
\begin{equation}
\frac{dN_{AB}(y)}{dy}\equiv\rho_{AB}(y)=w_{B}^{(c)}F_{B}(y)+w_{A}^{(c)}%
F_{A}(y), \label{w}%
\end{equation}
where $w_{A,B}^{(c)}$ is the number of wounded constituents (quarks and
diquarks) in the nucleus $A(B)$, whereas $F_{A}$ and $F_{B}$ are particle
densities emitted by one wounded quark or diquark in the nucleus $A$ and $B$,
respectively. Evaluation of $w^{(c)}$ for various processes considered in this
paper is given in the Appendix.

In the c.m. system we have, of course
\begin{equation}
F_{B}(y)=F_{A}(-y)\equiv F(y), \label{sym}%
\end{equation}
where from now on, $y$ shall always denote rapidity in the c.m. frame.

As explained in introduction, to extend the description to the region close to
the maximal and minimal rapidities, it is necessary to take into account at
least two other contributions to particle production. Taking this into
account, the Eq. (\ref{w}) is modified into
\begin{align}
\rho_{AB}(y)  &  =w_{B}^{(c)}F(y)+w_{A}^{(c)}F(-y)\nonumber\\
&  +\bar{w}_{B}^{(c)}U(y)+\bar{w}_{A}^{(c)}U(-y)+C_{B}(y)+C_{A}(-y),
\label{wnu}%
\end{align}
where $U(y)\equiv$ $U_{B}(y)=U_{A}(-y)$ represents the contribution from the
decay of one unwounded constituent (but belonging to a wounded nucleon) and
$\bar{w}_{A,B}^{(c)}$ is the number of unwounded constituents in the nucleus
$A(B)$. It can be simply determined using our assumption that each nucleon
consists of two active constituents (a quark and a diquark)%
\begin{equation}
w_{A,B}^{(c)}+\bar{w}_{A,B}^{(c)}=2w_{A,B}, \label{suma}%
\end{equation}
where $w_{A,B}$ is the number of wounded nucleons in the nucleus $A(B)$.

The terms $C_{A}$ and $C_{B}$ represent the ''intranuclear cascade'', i.e.,
the contribution from secondary interactions of particles created inside the nucleus.

The contributions $U_{A}$ and $C_{A}$ are expected to be significant only at
$y$ close to $-Y$ (the $c.m.$ rapidity of the nucleon in nucleus $A$).
Similarly the $U_{B}$ and $C_{B}$ are expected to be significant only at $y$
close to $Y$.

In nucleon-nucleon collisions, the ''cascade'' contribution is absent and we
obtain%
\begin{equation}
\rho_{NN}(y)=w_{N}^{(c)}\left[  F(y)+F(-y)\right]  +\bar{w}_{N}^{(c)}%
[U(y)+U(-y)], \label{pp}%
\end{equation}
where from the analysis of elastic $pp$ data we have found \cite{bb-AA} that
$w_{N}^{(c)}=1.187$. Following (\ref{suma}) we obtain $\bar{w}_{N}^{(c)}=0.813.$

The two equations (\ref{wnu}) and (\ref{pp}) describe the relation between
particle production in nucleon-nucleon, nucleon-nucleus and nucleus-nucleus
collisions implied by the model. They describe the data in terms of three
unknown functions. Since these functions have a well defined physical meaning,
the important result of our investigation is not only to verify the validity
of (\ref{wnu}) and (\ref{pp}) but also determination of the emission function
of a wounded constituent $F(y)$, the decay distribution of an unwounded
constituent $U(y)$ and the ''cascade'' contribution $C(y)$.

\section{d-Au collision}

The PHOBOS data on charged particle production in $d-Au$ collisions at
$\sqrt{s}=200$ GeV \cite{phob-dAu} cover the c.m. pseudorapidity range
$|\eta|\leq5.3$.

\begin{figure}[h]
\begin{center}
\includegraphics[scale=1.3]{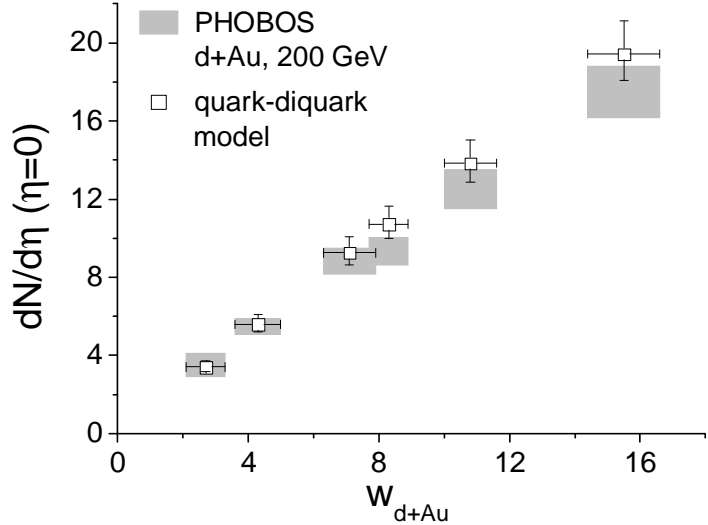}
\end{center}
\caption{Wounded quark-diquark model compared with the midrapidity $d-Au$ data
from PHOBOS coll.}%
\label{fig_dAu_mid}%
\end{figure}

The first step is to check if the data are consistent with the model at
$\eta=0$, where the simple formula [implied by (\ref{w}) and (\ref{pp})]
\begin{equation}
\rho_{d-Au}(0)=\rho_{NN}(0)\frac{w_{Au}^{(c)}+w_{d}^{(c)}}{2w_{N}^{(c)}},
\label{dAu_mid}%
\end{equation}
holds with $\rho_{NN}(0)$ being the density of particles produced in a single
$pp$ collision at $\eta=0$. In the further calculations we take $\rho
_{NN}(0)=2.31_{-0.16}^{+0.2}$ \cite{phob-pp}. In Fig. \ref{fig_dAu_mid}
$dN_{d-Au}/d\eta$ at $\eta=0$ is plotted versus $w_{d+Au}=w_{Au}+w_{d}$
(number of wounded nucleons in both colliding nuclei) and compared with the
model prediction (\ref{dAu_mid}). One sees that, within substantial
experimental errors, the data are consistent with the model\footnote{The
errors in model prediction reflect the inaccuracy in the $pp$ data i.e.
$\rho_{NN}(0)$.}. It is also seen, however, that the data for most central
collisions are somewhat below the results given by (\ref{dAu_mid}%
).

To investigate the rapidity region outside $y=0$, we first ignore the
''intranuclear cascade'' contributions $C_{d}$ and $C_{Au}$. This
simplification allows to investigate the data in terms of only two functions:
$F(y)\equiv F_{d}(y)=F_{Au}(-y)$ and $U(y)\equiv U_{d}(y)=U_{Au}(-y)$, being
the densities of particles produced from one wounded or unwounded constituent,
respectively. This condition is well justified for the deuteron since the
intranuclear cascade in this case is negligible. On the other hand, we expect
a discrepancy in the $Au$ fragmentation region. This discrepancy provides a
measure of the effect.

We construct $F(\eta)$ and $U(\eta)$ using the PHOBOS $d-Au$ data
\cite{phob-dAu}. First, to avoid propagation of the slight difference between
the model and the data observed at $\eta=0$ (c.f. Fig. \ref{fig_dAu_mid}), we
adjust data to the model at this point, multiplying at each rapidity by a
constant (for each centrality) factor $\rho_{d-Au}^{th}(0)/\rho_{d-Au}%
^{exp}(0)$. This allows to discuss the shape of the distributions in rapidity
independently of the absolute normalization (which is also consistent with the
model, as seen in Fig. \ref{fig_dAu_mid}).

Next we verified that in the pseudorapidity range $|\eta|\leq3.5$ the
contribution from the decay of unwounded constituents is not necessary for
description of data and thus one can take $U(\eta)=0$. It means that the data
in this region can be described solely in terms of wounded constituents. This
was confirmed by analysing the data on the deuteron side ($0\leq\eta\leq5.3$)
in terms of three functions: $F_{d}(\eta)=F(\eta)$, $F_{Au}(\eta)=F(-\eta)$
and $U_{d}(\eta)=U(\eta)$. From this analysis we have also found that
$F_{Au}(\eta)=0$ for $\eta>3.5$.

Using these findings and the $20-40\%$ centrality PHOBOS data\footnote{For the
$\eta<0$ one observes a certain difference between the STAR and PHOBOS data
for $0-20\%$ centrality. Therefore we preferred to take the $20-40\%$
centrality where this difference is much smaller.} it was possible to
determine $F(\eta)$ for $|\eta|\leq3.5$. For $\eta>3.5$ we used the $20-40\%$
and $40-60\%$ centrality PHOBOS data to obtain both $F(\eta)$ and $U(\eta)$.

\begin{figure}[h]
\begin{center}
\includegraphics[scale=1.3]{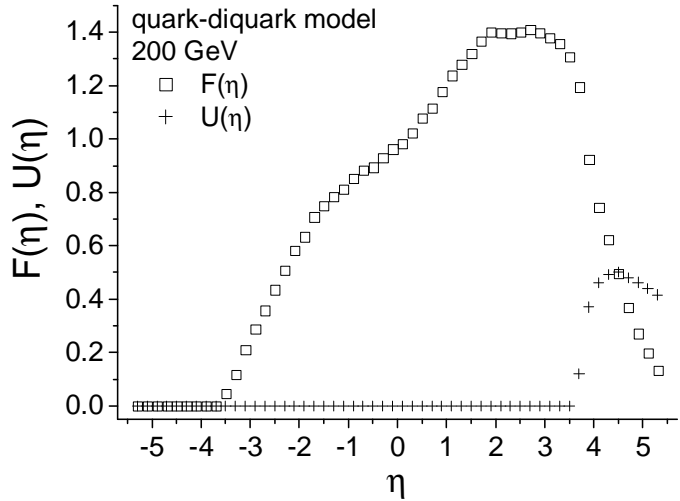}
\end{center}
\caption{{}The densities of particles produced from one constituent at
$\sqrt{s}=200$ GeV. Squares: wounded, $F(\eta)$; Crosses: unwounded, $U(\eta
)$.}%
\label{fig_uf}%
\end{figure}

In Fig. \ref{fig_uf} the functions $F(\eta)$ and $U(\eta)$ are shown. One sees
that, as expected \cite{bb,dt,bcl}, $U(\eta)$ is confined to the region close
to the maximal rapidity. One also observes that $F(\eta)$ is negligible for
$\eta\leq-3.7$.

\begin{figure}[h]
\begin{center}
\includegraphics[scale=1.5]{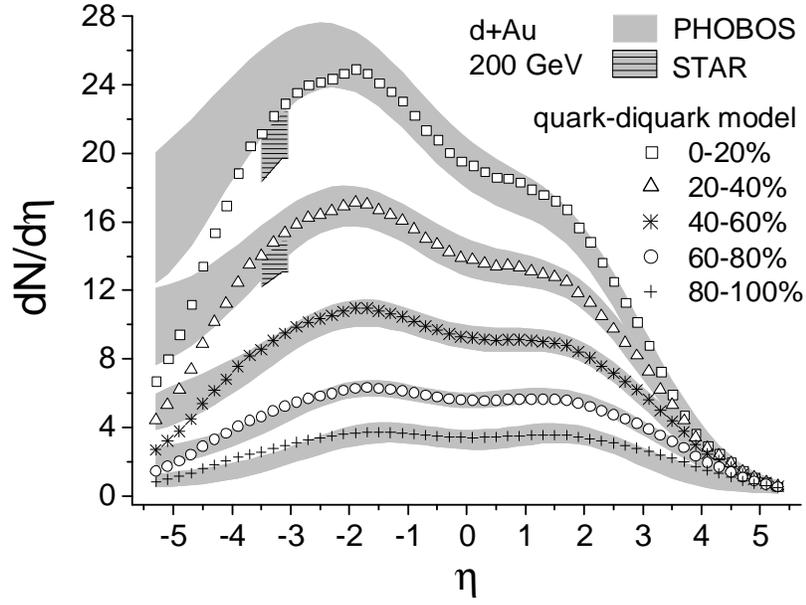}
\end{center}
\caption{Wounded quark-diquark model compared to $d-Au$ data from PHOBOS and
STAR coll. The data are normalized to the model prediction at $\eta=0.$ One
sees a generally good agreement except in the $Au$ fragmentation region where
the contribution from secondary interactions inside the target apparently
becomes important. }%
\label{fig_dAu}%
\end{figure}

Using these $F(\eta)$ and $U(\eta)$ one could evaluate the predictions of the
model for all centralities and in the full rapidity range. The results are
compared to the PHOBOS \cite{phob-dAu} and STAR \cite{star} data in Fig.
\ref{fig_dAu}. One sees that the agreement is satisfactory for $\eta\geq-4$.
The discrepancy for $\eta<-4$ can be attributed to the additional ''cascade''
contribution which is expected in the $Au$ fragmentation region. Large errors
do not allow, unfortunately, to perform a more quantitative analysis of this phenomenon.

All in all, the results presented in Fig. \ref{fig_dAu} show that the model
describes correctly the data.

\begin{figure}[h!]
\begin{center}
\includegraphics[scale=1.3]{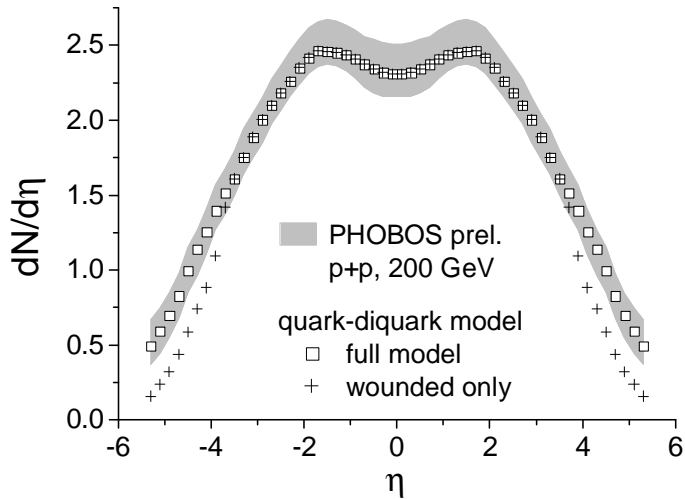}
\end{center}
\caption{Wounded quark-diquark model compared with the preliminary $pp$ PHOBOS
data. For comparison we also present the prediction without contribution from
unwounded constituents. One sees that it is necessary to account for the
data.}%
\label{fig_pp}%
\end{figure}

Since the functions $F$ and $U$ are now determined, it is possible to
construct the model prediction for $pp$ collisions. In Fig. \ref{fig_pp} this
is confronted with preliminary $pp$ PHOBOS data \cite{phob-pp}. We see that
the agreement is very good. For comparison we also present the prediction of
the model without contribution from unwounded constituents. It appears that
this contribution is indeed needed.

We feel that this result indicates that the physics of particle production in
$pp$ and $d-Au$ collisions is basically the same (apart from the trivial
difference in the form of intranuclear cascade).

It may be interesting at this point to speculate about the possible mechanism 
which may lead to the results for $F(\eta)$ and $U(\eta)$ shown in Fig. \ref{fig_uf}. 
The very broad rapidity distribution of particles emitted by the wounded 
source suggests a colour-exchange mechanism. The simplest possibility is the multi-gluon 
exchange between the projectile and target (every gluon exchange changes the colour of the source and thus may lead to particle production). Since every constituent consists of many partons and since the probability of exchanging a gluon between two partons does not depend on their rapidity difference (gluon is a particle of spin 1), one may fairly easily obtain a rather broad rapidity distribution of the produced particles. As discussed in \cite{bj,bc} this picture can naturally accommodate the asymmetric shape of $F(\eta)$ (the shoulder seen at $\eta=-1$ is -most likely- the kinematic effect related to the transformation from rapidity to pseudorapidity). On the other hand, particle production from the unwounded source can be understood as resulting from the decay of a string spanned between the wounded and unwounded constituent. The details of this picture and its derivation from fundamental theory represents an interesting problem which, however, goes beyond the scope of the present investigation.

\section{Au-Au and Cu-Cu collisions}

\begin{figure}[h]
\begin{center}
\includegraphics[scale=1.3]{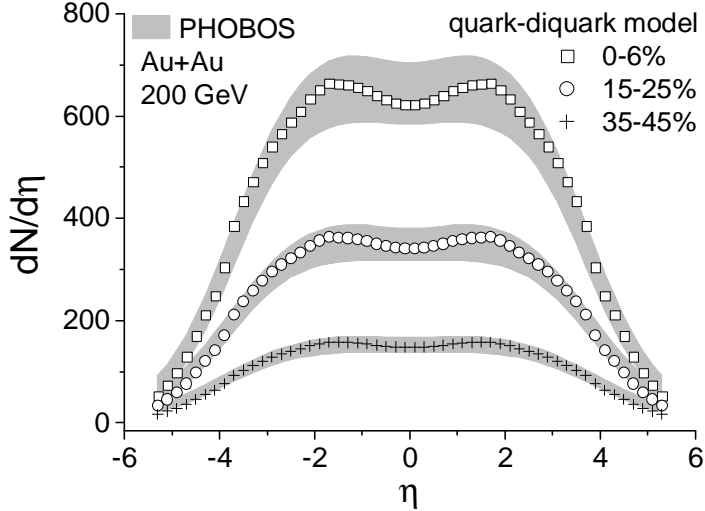}
\end{center}
\caption{Wounded quark-diquark model compared with the $Au-Au$ data from
PHOBOS coll.}%
\label{fig_auau}%
\end{figure}

In Fig. \ref{fig_auau} we compare the predictions of the wounded quark-diquark
model with the PHOBOS data \cite{phob-AuAu} in $Au-Au$ collisions. This
observation shows that the intranuclear cascade is not effective for central
nucleus-nucleus collisions\footnote{A closer look shows that there is some
indication of discrepancy in the fragmentation region in case of most
peripheral collisions.}, thus suggesting that the secondary interactions in
the nucleus lead to particle production mainly if they happen on the spectator
nucleons. It would be interesting to investigate this point in more detail
when more precise data are available.

The values of the numbers of wounded constituents for various centralities are
given in the Appendix.

We have found in the previous section that $F(\eta)$, the contribution form
wounded constituents, largely dominates the spectrum in the region $|\eta
|\leq4.5$. It then follows from (\ref{wnu}) that, for symmetric
nucleus-nucleus collisions, the ratio of particle densities produced by
various nuclei must be a constant, independent of $\eta$. Indeed, we have
\begin{equation}
R_{Au/Cu}(\eta)\equiv\frac{\rho_{AuAu}(\eta)}{\rho_{CuCu}(\eta)}%
=\frac{w_{Au}^{(c)}}{w_{Cu}^{(c)}}=R_{Au/Cu}(0).
\end{equation}
This consequence of the model is known to be very well satisfied by $Au-Au$
and $Cu-Cu$ data \cite{phob-cucu}.

\begin{figure}[h]
\begin{center}
\includegraphics[scale=1.3]{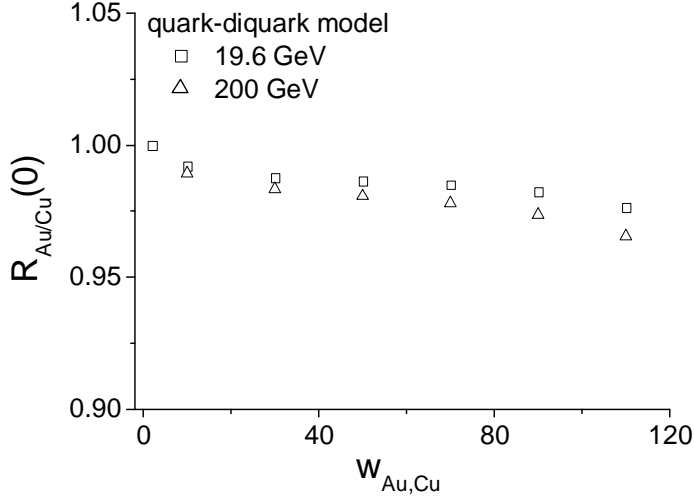}
\end{center}
\caption{The ratio of density of particles produced at midrapidity in $Au-Au$
collisions to that produced in $Cu-Cu$ collisions at the same number of
wounded nucleons. For comparison we also present the results at $19.6$ GeV
c.m. }%
\label{fig_ratio}%
\end{figure}

In Fig. \ref{fig_ratio} $R_{Au/Cu}(0)$ evaluated from the model is shown
versus the number of wounded nucleons (the same for $Au$ and $Cu$). One sees
that it is very close to $1$, also in agreement with data \cite{phob-cucu}%
.

\section{Conclusions and comments}

Our conclusions can be formulated as follows.

(i) We have compared the wounded quark-diquark model with the PHOBOS data on
particle production in $p-p,$ $d-Au,$ $Cu-Cu$ and $Au-Au$ collisions at $200$
GeV c.m. energy in the full range of rapidity. The shape of the rapidity
distribution is well reconstructed. The overall normalization of the $d-Au$
data is not described so well, although it is consistent with the model within
the (still rather large) experimental uncertainties.

(ii) The particle emission function from one wounded constituent is
determined. Its most important features are: (a) a large maximum in the
forward direction; (b) a significant fraction of particles emitted in the
backward hemisphere; (c) strong suppression of particle emission in the target
fragmentation region.

(iii) Particle emission from unwounded constituents contributes only to the
fragmentation region of the projectile.

(iv) Particle production from secondary interactions inside the nucleus are
confined to its fragmentation region, in agreement with theoretical
expectations. These secondary interactions seem to have stronger effect on
particle production in nucleon-nucleus than in the central nucleus-nucleus collisions.

(v) For the good agreement of the model with data it was crucial to accept that each wounded constituent emits the same density of particles. This somewhat surprising result indicates that intensity of particle emission is mainly determined by the colour content of the source (the colour content of quark and diquark are the same). This suggests the colour-exchange models as a possible mechanism responsible for particle production, in harmony with the discussion at the end of Section 3.

Following comments are in order.

(i) Contrary to the common prejudice, good agreement of the wounded
constituent model with $Au-Au$ and $Cu-Cu$ data is \textit{not} in
contradiction with the collective phenomena observed in heavy ion collisions.
It only shows that in all hadronic collisions \textit{the early stage} of the
particle production process can be understood as a superposition of
contributions from hadronic constituents. This does not preclude further
collective evolution of the system which obviously must be more visible in the
system produced in collision of two heavy nuclei than, e.g., in
nucleon-nucleon collision.

(ii) The wounded constituent picture has, however, important consequences for
the properties of the hot matter created in heavy ion collisions. First of
all, it implies that most of the entropy must be produced already at the very
early stage of the collision. It also implies rather early equilibration in transverse direction, presumably already at the level of nucleon-nucleon collisions \cite{bec,b1,raf}. Finally, the absence of the visible effects of the longitudinal pressure in $Au-Au$ collisions (which would modify somewhat the rapidity distribution) suggests that the early evolution of the plasma may be dominated by the purely {\it transverse} hydrodynamic expansion while the longitudinal evolution is described by free-streaming. This possibility (which solves the "problem of early equilibration") was investigated recently \cite{bcf} and shown to be in agreement with data on elliptic flow. One may thus consider our result as an indirect confirmation of the hypothesis formulated in \cite{bcf}. 

(iii) The model describes particle production in nucleon-nucleon collisions as
emission from two (left- and right-moving) sources, both populating most of the available phase-space. This picture suggests presence of specific long-range, forward-backward correlations. It would be interesting to study these correlations in more detail.

(iv) In the present investigation we have entirely neglected the possible
dependence on transverse mass, thus implicitly assuming that $m_{t}$
distributions do not vary significantly with the varying centrality of the
collision. This is approximately correct for the bulk of created particles
(mostly pions at small transverse momentum) but it would be certainly
interesting to investigate limits of the wounded constituent model at higher
transverse momenta and for heavy particles.

(v) The model provides a natural qualitative explanation of the ''stopping''
of a high-energy nucleon in a collision with the heavy nucleus \cite{bg}.
Indeed, since the contribution from an unwounded constituent dominates the
very end of rapidity phase-space and since the number of unwounded
constituents in the nucleon passing through a nucleus is smaller than that in
a nucleon-nucleon collision, one expects less high-energy nucleons in the
former case. A quantitative estimate of this effect may provide information
about the momentum distribution of the constituents (quark and diquark) inside
the nucleon.

\bigskip

\textbf{Acknowledgements}

We would like to thank W. Busza and R. Holynski for very useful discussions
about the data, and to K. Fialkowski and K. Zalewski for the critical reading
of the manuscript. A comment by W. A. Zajc on nuclear stopping power, and by
H. Bialkowska and B. Boimska on transverse momentum dependence are highly
appreciated. This investigation was partly supported by the MEiN research
grant 1 P03B 045 29 (2005-2008).

\appendix                        

\section{Appendix: Estimate of the number of wounded constituents}

\textbf{A1. Au-Au}

Evaluation of the number of wounded quarks and diquarks $w_{Au}^{(c)}$ in
$Au-Au$ collision is described in detail in \cite{bb-AA}. For the three
centralities, $0-6\%,$ $15-25\%$ and $35-45\%$, presented in Fig.
\ref{fig_auau} we obtain $w_{Au}^{(c)}=320$, $175$ and $76$, respectively. The
number of unwounded constituents is determined from (\ref{suma}) with the
values $2w_{Au}$ taken from \cite{phob-AuAu} ($2w_{Au}=344,$ $200$ and $93$,
respectively). The same procedure is used for $Cu-Cu$ collisions.

\noindent\textbf{A2. d-Au}

Evaluation of the number of wounded quarks and diquarks in $d-Au$ system is
somewhat more complicated.

First we assume that each wounded nucleon in $Au$ is hit only
once\footnote{This is a good approximation since there is no significant
difference between the number of wounded nucleons in $Au$ and the number of
collisions \cite{phob-dAu}.}, so the number of wounded constituents in $Au$ is
given by: $w_{Au}^{(c)}=w_{Au}w_{N}^{(c)}$, where $w_{Au}$ is the number of
wounded nucleons in $Au$ (provided by the PHOBOS coll. \cite{phob-dAu}) and
$w_{N}^{(c)}=1.187$ \cite{bb-AA}.

To estimate the number of wounded quarks and diquarks in the deuteron let us
first consider the nucleon-nucleus collision.

The average number of wounded quarks $w_{N,k}^{(q)}$ in the nucleon which
underwent $k$ inelastic collisions is given by the straightforward counting of
probabilities
\begin{equation}
w_{N,k}^{(q)}=1-(1-p_{q})^{k},
\end{equation}
where $p_{q}$ is the probability for a quark to interact in a single $pp$
collision. In \cite{bb-AA} we have found that $p_{q}\approx w_{N}%
^{(c)}/3=0.395$. One can write analogous formula for $w_{N,k}^{(diq)}$ i.e.
number of wounded diquarks in the nucleon which underwent $k$ inelastic collisions.

Thus, the number of wounded constituents (quarks and diquarks) in the nucleon
which underwent $k$ inelastic collisions is
\begin{equation}
w_{N,k}^{(c)}=2-(1-p_{q})^{k}-(1-p_{diq})^{k}, \label{k2}%
\end{equation}
where $p_{diq}$ is the probability for a diquark to interact in a single $pp$
collision and $p_{diq}\approx2p_{q}=0.79$.

For the average number $<k>$ of collisions per one wounded nucleon in deuteron
we take $<k>=n_{coll}/w_{d}$, using the values of $n_{coll}$ (number of
collisions) and $w_{d}$ (number of wounded nucleons in deuteron) given by the
PHOBOS coll. \cite{phob-dAu}.

Replacing in (\ref{k2}) $k$ by $<k>$ we obtain\footnote{We have verified that
for the Glauber distribution the error induced by replacing $k$ by $<k>$ does
not change the results significantly.} $w_{N,<k>}^{(c)}$ and, finally, the
number of wounded constituents $w_{d}^{(c)}$ in deuteron for a given
$n_{coll}$ and $w_{d}$ reads%
\begin{equation}
w_{d}^{(c)}=w_{d}w_{N,<k>}^{(c)}.
\end{equation}

This procedure gives: $w_{d}^{(c)}=3.95,$ $3.66,$ $3.1,$ $2.3,$ $1.6$ for
$0-20\%,$ $20-40\%,40-60\%,$ $60-80\%$ and $80-100\%$ centrality, respectively.

\end{document}